\pdfoutput=1

\documentclass[twocolumn,showpacs,superscriptaddress,groupedaddress,floatfix]{revtex4-1}

\usepackage{color}
\usepackage[]{graphicx} 
\usepackage{bm}       
\usepackage{amssymb}
\usepackage{amsmath}  
\usepackage{float}
\usepackage[english]{babel}

\def\perkg{\mathrm{kg^{-1}}}
\def\per2m{\mathrm{m^{-2}}}
\def\persec{\mathrm{s^{-1}}}
\def\perTesla{\mathrm{T^{-1}}}

\def\perm{\mathrm{m^{-1}}}

\def\2D{\mathrm{2D}}
\def\3D{\mathrm{3D}}
\def\kB{k_\mathrm{B}}

\def\N2{N$_2$}
\def\oz{\omega_{0}}

\begin{document}

\title{Optical and magnetic measurements of gyroscopically stabilized graphene nanoplatelets levitated in an ion trap}
\author{Pavel Nagornykh}
\thanks{These authors contributed equally to this work. P.N. is currently at Center for Nonlinear Dynamics, Department of Physics, University of Texas, Austin, TX.}
\author{Joyce E. Coppock}
\thanks{These authors contributed equally to this work. P.N. is currently at Center for Nonlinear Dynamics, Department of Physics, University of Texas, Austin, TX.}
\author{Jacob P. J. Murphy}
\affiliation{Department of Physics, University of Maryland, College Park, MD}
\author{B. E. Kane}
\email{bekane@umd.edu}
\affiliation{Joint Quantum Institute, University of Maryland, College Park, MD}
\affiliation{Laboratory for Physical Sciences, University of Maryland, College Park, MD}
\date{\today}
\begin{abstract}
Using optical measurements, we demonstrate that the rotation of micron-scale graphene nanoplatelets levitated in a quadrupole ion trap in high vacuum can be frequency locked to an applied radio frequency electric field $\bm{E}_{\mathrm{rf}}$.  Over time, frequency locking stabilizes the nanoplatelet so that its axis of rotation is normal to the nanoplatelet and perpendicular to $\bm{E}_{\mathrm{rf}}$. We observe that residual slow dynamics of the direction of the axis of rotation in the plane normal to $\bm{E}_{\mathrm{rf}}$ are determined by an applied magnetic field. We present a simple model that accurately describes our observations. From our data and model we can infer both a diamagnetic polarizability and a magnetic moment proportional to the frequency of rotation, which we compare to theoretical values. Our results establish that trapping technologies have applications for materials measurements at the nanoscale.

\end{abstract}

\pacs{}
\maketitle

\section{Introduction} \label{sec:introduction}
The field of trapping, stabilization, and cooling of nanoscale particles has progressed rapidly in recent years, and this technology has great promise to make contributions in fields ranging from
studying quantum behavior of macroscopic objects\citep{Goldwater2016}\citep{Jain2016} 
  to developing highly sensitive measurements, 
both of the properties of the trapped particles themselves\citep{Morago2013}\citep{Spesyvtseva2016}\citep{Howder2015} 
and of weak forces acting on the particles\citep{Hoang2016b}\citep{Ranjit2016}.
A recent advance in this field, enabled by the development of traps in good vacuum chambers, is the ability to impart rotation to particles at frequencies in excess of 1 MHz using angular momentum provided by circularly polarized light\cite{Kane2010}\citep{Arita2013}\citep{Kuhn2015}\citep{Kuhn2016}.
Rapid rotation could lead to substantial improvements in measurement sensitivity, since the particle orientation is stabilized in a manner similar to a classical gyroscope.  Particles set into rotational motion purely by circularly polarized light, however, suffer from the deficiency that they are free running oscillators, with the ultimate frequency determined by frictional torques.  
To improve measurements on rotating particles, it is desirable to lock the rotation frequency to an external oscillator. 

Using a quadrupole ion trap and parametric feedback to stabilize trapped particles in high vacuum\citep{Gieseler2012}\citep{Nagornykh2015}, 
we have succeeded in locking the rapid ($\sim$20 MHz) rotation of charged $\mu$m-scale graphene nanoplatelets to an externally applied radio frequency electric field, $\bm{E}_{\mathrm{rf}}$, which couples to the nanoplatelet via a permanent electric dipole moment $\bm{p}$ that is inevitably present on the charged, irregularly shaped platelets.  Once locked, the nanoplatelets exhibit a remarkable gyroscopic stabilization of their orientation, and the slow ($\sim$1 Hz) residual dynamics of their orientation are determined by their interactions with the local dc magnetic field $\bm{B}$.  From measurements of the slow dynamics, we have observed a large diamagnetic response that is characteristic of graphite and hypothesized for graphene\citep{Koshino2009}\citep{Ominato2013}. We also observe a magnetic dipole moment associated with the rapid rotation of the charged platelet and can measure $g$ factors of the levitated nanoplatelet.

Our measurements imply a magnetic moment sensitivity of order $10^{-20}~\mathrm{J}\,\perTesla$ and a torque sensitivity of order $10^{-22}$ J.  While our measurements are on rather poorly characterized samples (they are multilayer with an uncertain number of layers), the technique holds great promise for future measurements of the magnetic properties of single layer graphene and other two dimensional materials, a subject of significant theoretical research\citep{Principi2009}\citep{Gutierrez2016}\citep{Hesse2014}, 
or indeed of any $\mu$m-scale material with an anisotropic magnetic response. 

\section{Experimental Apparatus} \label{sec:apparatus}

\begin{figure*}
\begin{center}
\includegraphics[width=\textwidth]{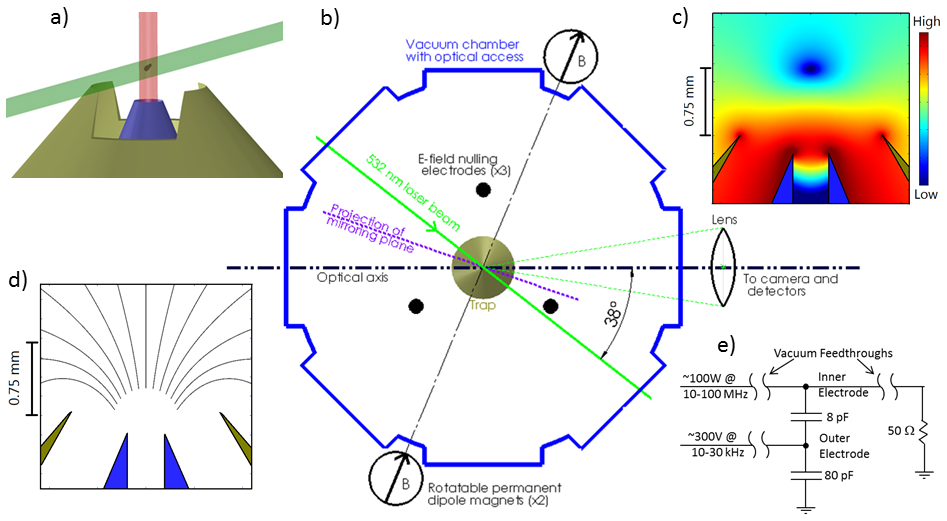}
\end{center}
\caption{Experimental apparatus. (a) The trap is comprised of two coaxial electrodes, an inner electrode that is hollow for transmission of an axial laser beam and an outer electrode engineered with a slot to break rotational symmetry.  (b) The trap is situated at the center of an octagonal vacuum chamber with a 532 nm laser illuminating the particle at a $38^{\circ}$ angle from the axis of the collection optics.  E-field nulling electrodes are inside the chamber and adjustable orientation permanent dipole magnets are outside the chamber. Cross-sectional plots of ponderomotive potential and rf electric field lines are plotted in (c) and (d) respectively.  Both sections lie in the plane that is vertical and perpendicular to the slot in the tip of the outer electrode.  (e) Lumped circuit diagram of the trap electrodes and the rf termination located outside the vacuum chamber.}
\label{fig:apparatus}
\end{figure*}

The experimental apparatus was developed explicitly to operate in the high vacuum environments necessary for high frequency rotation measurements (Fig.\:\ref{fig:apparatus}) and is described more fully in Ref.\:\citep{Nagornykh2015}.  The trap consists of two coaxial conical electrodes and is designed to have three easily distinguished translational frequencies so that parametric feedback can be used to stabilize translational motion of the trapped particles in high vacuum\citep{Nagornykh2015}. For the primary particle under study in this article, the frequencies were $\sim$300, $\sim$450, and $\sim$750 Hz; typical frequencies range from approximately three-quarters to two times these values. The eigenfrequency of motion parallel to the central axis of the trap is given by

\begin{align} 
\omega_{z} =\frac{1}{\sqrt{2}}\frac{q}{m}\frac{V_{out}}{\Omega_{t}z_{0}^{2}}, \label{eq:eigenf} 
\end{align}
where $q$ is the charge on the particle, $m$ is its mass, $V_{out}$ is the amplitude of the voltage applied to the outer electrode, $\Omega_{t}$ is the frequency of $V_{out}$, and $z_0$ is a parameter determined by modeling the electrode configuration\citep{Coppock2017}. For the present experiment, $V_{out} \approx 300$ V, $\Omega_{t}/\left(2\pi\right) \approx 15$ kHz, and $z_0 = 1.7$ mm. 

Because the process of introducing a nanoplatelet to the trap generates water vapor, particle collection is performed in an antechamber that is separated by a gate valve from the high vacuum chamber in which experiments are performed.  After being generated by liquid exfoliation\citep{Nicolosi2013} and introduced into the collection chamber at $\sim$500 mTorr by an electrospray source, which imparts a positive charge to the surface of the flake, several particles are collected in a trap mounted on a linear translation stage and all but one are expelled from the trap. Particles selected for study in these experiments have mass of order $10^{-17}$~kg and typical charge of order $+1000|e|$.  While the variability of sample size is substantial, flakes typically have lateral sizes of order 1 $\mu$m, with thickness of order 10 layers\citep{Hernandez2008}.  

After a particle is selected, the collection chamber is pumped to $\sim$10$^{-2}$ Torr and the gate valve into the high vacuum chamber is opened.  The trap containing the particle is transported into the high vacuum chamber on the translation stage and brought into close proximity to the high vacuum trap.  It is subsequently transferred to the high vacuum trap by gradual adjustment of the two trap voltages.  The collection and exchange process is described in detail in Ref. \citep{Coppock2017}.  Prior to pumping down to the experimental pressures ($<10^{-7}$ Torr), stray dc electric fields are nulled using electrodes near the trap\citep{Eltony2013}\citep{Nagornykh2015th}.  In the present experiments, parametric feedback is used to stabilize the particle's translational motion, but the temperature associated with translational motion remains near 300 K. 

Optical measurements are made using the light from 532 nm laser impinging perpendicular to the trap central axis.  Scattered light is focused by a lens onto a CCD camera and onto dual high speed single photon counters\footnote{Hamamatsu MPPC Module C13366-1350GD. The $\sim$20 ns pulses output by this unit are shortened to $\sim$1 ns to increase frequency response to $>$500 MHz.}. While the difference in photon counter signals is used for stabilization of translational motion, the sum signal is used for all measurements presented below.  Maximum power density of the laser at the trapped particle is typically around 0.2 $\mathrm{W}\,\mathrm{cm}^{-2}$.  Above this power particles rapidly discharge in high vacuum\citep{Nagornykh2015}, while below it they can exhibit negligible discharging over periods of weeks.

For all data presented, the axis of polarization of the 532 nm laser light is linear and parallel to the trap axis.  After high vacuum conditions are established, rotation is imparted to the trapped nanoplatelet using circularly polarized light from a 671 nm laser (operated at similar power densities as the 532 nm laser) that propagates along the trap axis (Fig.\,\ref{fig:apparatus}a).  After a period of illumination from the circularly polarized laser, a sharp peak develops in the power spectral density (PSD) of the summed detector signals (Fig.\,2), indicating that the nanoplatelet is rapidly rotating.  The time taken for the rotation peak to appear varies widely between samples, from a few minutes to an hour, which is expected given the variability in nanoplatelet area and thickness.

Frequency locking of the rotation is achieved by applying a large oscillating electric field on the inner electrode of the ion trap (Fig. \ref{fig:apparatus}(d,e)).  The rf is sourced by a 100 W power amplifier, is routed into and out of the vacuum chamber on separate feedthroughs, and is terminated by 50 $\Omega$.  Resonances and reflections of the rf are minimized so that the rf frequency can be swept over a large range with minimal electric field amplitude variations at the location of the trapped particle.  Modeling of the trap indicates that the amplitude of the rf electric field at the trap center reaches 20 $\mathrm{kV}\,\perm$ at maximum power\citep{Coppock2017}.  Note that for the typical charge to mass ratios of trapped particles ($q/m\cong 10~\mathrm{C}\,\perkg$), the maximum rf field at 10 MHz leads to particle motion of order 0.1 nm, which is far smaller than the thermal spreading of the particle in the trap.

\begin{figure}
\begin{center}
\includegraphics[scale=1,draft=false]{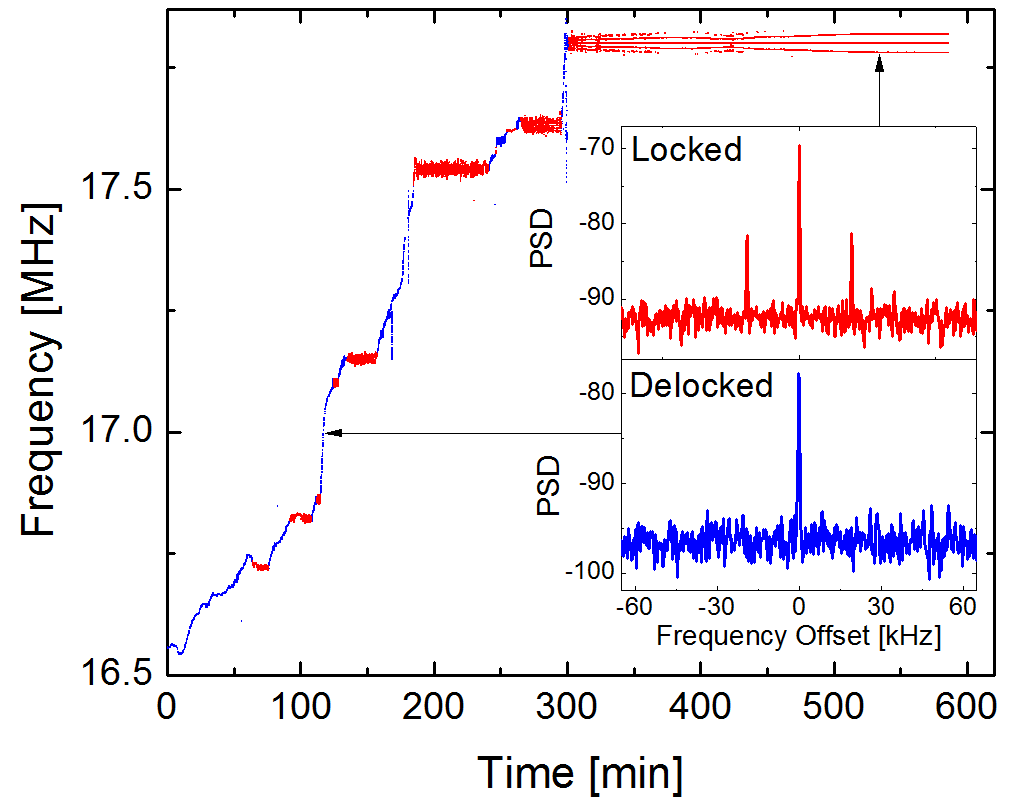} 
\end{center}
\caption{Locking the rotation of the flake to an external rf electric field.  Blue regions indicate where the particle is only experiencing torques from the circularly polarized laser, which spin the flake up to MHz frequencies.  In red regions $\bm{E}_{\mathrm{rf}}$ is turned on and the flake becomes locked to the applied field.  Insets show that sidebands appear on the optical signal only when locking occurs.}
\label{fig:locking}
\end{figure}

Locking is established by applying an rf signal at 
\textit{one half}
the frequency (typically 10-20 MHz) of the peak observed in the PSD of the optical signal (Fig.\:\ref{fig:locking}).  (The factor of $1/2$ is explained in Sec.\:\ref{subsec:optscattering}.)  When locked, the PSD peak remains at a constant frequency and sidebands appear in the spectrum, located at a distance of order 10 kHz from the central peak.  The particle remains locked when the source of circularly polarized light is turned off, and the offset frequencies of the sidebands stabilize.  
Stable locking has been achieved for periods of several weeks on a few individual particles in our laboratory.

\section{Model of spinning nanoplatelet locked to an external electric field} \label{sec:dynamicsmodel}
To gain insight into locked flake behavior, we have developed a simple dynamical model with the following assumptions: 
(1) \textit{The flake is an irregularly shaped rigid two dimensional object}, i.e. its lateral dimensions greatly exceed its thickness.
(2) \textit{The flake possesses a permanent electric dipole moment $\bm{p}$ lying in the 2D plane of the flake.} $~\bm{p}$ arises because the charge and mass distributions are distinct in charged conducting objects and consequently any such object lacking symmetry will experience a torque in an electric field.
(3) \textit{The flake always rotates around an axis perpendicular to the flake plane.} This assumption is certainly \textit{not} valid in general; however, in systems where rotational energy exceeds  the thermal energy, $\kB T$, and where angular momentum is conserved, but where rotational energy can be rapidly exchanged with internal degrees of freedom,
rotation tends to stabilize around the principle axis with the largest value of the moment of inertia tensor\cite{Purcell1979}.  This axis for 2D objects is always perpendicular to the object's surface, with moment of inertia denoted by   
$I_{\perp}$.  In our system thermal rotation 
$\omega_{th}\cong \sqrt{\kB T/I_{\perp}}\cong10^{4}~\persec$, much less than the rotational angular velocities we measure when the rotation is locked to $\bm{E}_{\mathrm{rf}}$.  Also, we will focus on the low frequency response of the flake to external torques, and thus fluctuations in the instantaneous axis of rotation of the flake are likely to be averaged out.

\begin{figure}
\begin{center}
\includegraphics[scale=0.5,draft=false]{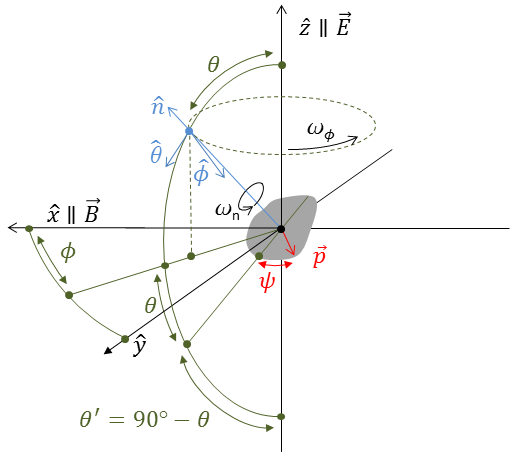} 
\end{center}
\caption{Diagram of a spinning graphene nanoplatelet.  We assume that the flake spins on an axis 
$\bm{\hat{n}} $ normal to its surface and that a permanent electric dipole moment $\bm{p}$ lies in the flake plane.  An oscillatory electric field is parallel to the z axis, and a dc magnetic field lies along the x axis.  For the calculations, the local flake frame Cartesian coordinate system denoted by 
$[\bm{\hat{\theta}},\bm{\hat{\phi}},\bm{\hat{n}}]$  is used, which does $not$ rotate with the particle.}
\label{fig:spinangles}
\end{figure}

In order to model flake dynamics, we assume an oscillatory electric field $\bm{E}_{\mathrm{rf}}$ is directed along 
$\bm{\hat{z}}$, and the components of $\bm{E}_{\mathrm{rf}}$ in the local flake frame (Fig.\:\ref{fig:spinangles}) are
\begin{align} \nonumber
E_{\theta}&=E_{0}\sin \theta \cos \omega_{0} t, \\ \nonumber
E_{\phi}&=0, \\ \nonumber
E_{n}&=-E_{0} \cos \theta \cos \omega_{0} t, \\ \nonumber
\end{align} 
where  $E_{0}$ and $\omega_{0}$ are the amplitude and angular velocity of $\bm{E}_{\mathrm{rf}}$.  The components of $\bm{p}$ are
\begin{align} \nonumber
p_{\theta}&=p_{0} \cos (\omega_{0} t +\psi ), \\ \nonumber
p_{\phi}&=p_{0} \sin   (\omega_{0} t +\psi), \\ \nonumber
p_{n}&=0, \\ \nonumber
\end{align} 
where $\psi$ is the phase deviation of the dipole from perfect synchrony with the applied field, and is assumed to vary slowly compared to $\omega_{0}$. We calculate the torque, 
$\bm{N} = \bm{p} \times \bm{E}_{\mathrm{rf}}$, on the flake:
\[
\bm{N} =p_{0}E_{0}
\begin{vmatrix}
\bm{\hat{\theta}} & \bm{\hat{\phi}} & \bm{\hat{n}}  \\ 
\cos (\oz t +\psi) & \sin   (\oz t +\psi) &0  \\
\sin \theta \cos \oz t & 0 & -\cos \theta \cos \oz t 
\end{vmatrix}.
\]
The results, derived by using sine and cosine formulas and by retaining only the terms that are not rapidly oscillating, are
\begin{align}
N_{\theta}&=-\frac{1}{2}p_{0}E_{0}\cos \theta \sin \psi, \label{a} \\ 
N_{\phi}&=+\frac{1}{2}p_{0}E_{0}\cos \theta \cos \psi, \label{b} \\ 
N_{n}&=-\frac{1}{2}p_{0}E_{0}\sin \theta \sin \psi. \label{c} \\ \nonumber
\end{align}

\subsection{Oscillations in rate of axial rotation} \label{subsec:oscaxialrot}
First, we calculate the in-plane oscillations of the dipole around its minimum energy orientation.  These are rotations around the axis $\bm{n}$ normal to the flake plane:
\[N_{n}=\dot{L}_{n}=I_{\perp}\dot{\omega}_{n}=I_{\perp}\ddot{\psi}=-\frac{1}{2}p_{0}E_{0}\sin \theta \sin \psi\]
where $\bm{L}$ is angular momentum of the flake.  For small $\psi$, this equation will yield simple harmonic oscillations with angular velocity:
\begin{equation} \label{d}
\omega_{\psi}^{2}=\frac{p_{0}E_{0}}{2 I_{\perp}} \sin \theta \equiv \Omega^{2} \sin \theta .
\end{equation}
These oscillations lead to deviations of $\omega_{n}$, the instantaneous angular rotation velocity of the flake, from $\omega_{0}$.  They are entirely analogous to the torsional oscillations observed recently in optically levitated non-spherical particles\citep{Hoang2016b} and manifest themselves as sidebands to the measured frequency of rotation observed in the PSD of the scattered light.  They are perhaps best termed as librations of the flake viewed in a frame rotating synchronously with $\bm{E}_{\mathrm{rf}}$.  When averaged over times longer than the period of the $\omega_\psi$ oscillations, $\omega_n=\omega_0$.  

\subsection{Optical scattering from nanoplatelets and the appearance of sidebands} \label{subsec:optscattering}
The nanoplatelets measured in the experiments are typically comparable in size or large compared to the 532 nm wavelength of the probing light, while the thickness of the flake is much smaller than the wavelength.  In the Rayleigh-Gans approximation\citep{Gordon2007}, where the particle has negligible effect on the incident radiation, the intensity of light scattering is dominated by the form factor:
\begin{equation}\label{e}
F (\bm{\tilde{k}})=\left| \int_{S} d\bm{r} e^{i \bm{\tilde{k}} \cdot \bm{r}} \right|^{2},
\end{equation}
where $\bm{\tilde{k}}$ is the wavevector difference between incident and scattered light and the integral is over the surface of the flake.  We neglect the polarization of the light, since our measurements are performed when the direction of the polarization is $\perp$ to $\bm{\tilde{k}}$. For a 2D object lying in the plane $\perp \bm{\hat{n}} $:
\[F(\tilde{k}_{\theta},\tilde{k}_{\phi},\tilde{k}_{n})=F(\tilde{k}_{\theta},\tilde{k}_{\phi},-\tilde{k}_{n}).\]
Additionally, time reversal symmetry implies that  $F (\bm{\tilde{k}})=F (-\bm{\tilde{k}}) $.  Thus:
\[F(\tilde{k}_{\theta},\tilde{k}_{\phi},\tilde{k}_{n})=F(-\tilde{k}_{\theta},-\tilde{k}_{\phi},\tilde{k}_{n}).\]
This result implies that the form factor is unchanged when the nanoplatelet is rotated $180^{\circ}$ around an axis perpendicular to its surface.  The nanoplatelet rotating at angular velocity $\omega_{n} $ around 
$\bm{\hat{n}} $  will thus only have optical modulation at even harmonics. This is, in fact, what we have observed in our experiments, although odd harmonics with significantly smaller magnitude are observable for some particles.

The optical signal detected from a rotating nanoplatelet is  
\[I(t)=\sum_{n} A_{n} e^{2 n i (\omega_{0} t + \psi_{0} \sin \omega_{\psi} t )},\]
where $\psi_0$ is the amplitude of the small in-plane oscillations discussed in Sec.\:\ref{subsec:oscaxialrot}. We sum over only even harmonics, as discussed above, and we explicitly include phase deviations from slow oscillations around the rotation axis.  We next make the approximation valid when $\psi_{0} $ is small:
\[e^{2 n i  \psi_{0} \sin \omega_{\psi} t } \cong 1+ 2 n i \psi_{0} \sin \omega_{\psi} t = 1 + n  \psi_{0} ( e^{i \omega_{\psi} t }-e^{-i \omega_{\psi} t }). \]
Thus:
\[I(t)=\sum_{n} A_{n} [e^{2 n i \omega_{0} t}+ n \psi_{0}( e^{i (2 n  \omega_{0}+\omega_{\psi} ) t}-  e^{i (2 n  \omega_{0}-\omega_{\psi} ) t})].\]
The sidebands are displaced from all harmonics by $\pm \omega_{\psi} $. The power in the sidebands relative to the harmonics can be used to determine $\psi_{0} $:
\begin{equation}\label{f}
\frac{P(2 n \omega_{0} \pm \omega_{\psi}) }{P(2 n \omega_{0} )} = n^{2} \psi_{0}^{2}.
\end{equation}

If rotational motion of the locked flake is thermal, then, from the equipartition theorem:
\begin{equation}\label{g}
k_{B} T \approx \frac{1}{2} p_{0} E_{0} \sin \theta \left\langle \psi ^{2} \right\rangle,
\end{equation}
where $\left\langle \psi ^{2} \right\rangle = \left(1/2\right) \psi_{0}^{2}$ is the average value of $\psi$.  Using $n=1$ because we study the signal at the second harmonic, Eqs.\:\ref{f} and \ref{g} simplify to:

\begin{equation}\label{eq:sidebandpower}
\frac{P(2 \omega_{0} \pm \omega_{\psi}) }{P(2 \omega_{0} )} = \frac{4 k_B T}{p_0 E_0 \sin \theta},
\end{equation}
which allows us to determine the electric dipole moment $p_0$ if the orientation angle $\theta$ is known.  It is likely that axial oscillations are rapidly thermalized with internal degrees of freedom of the flake and that $T \approx~$300 K.  Although substantial heating of the illuminated particle is possible in a high vacuum environment, the low laser powers used in the present experiments would not be expected to raise the internal temperature significantly above 300 K, even if the flake were a perfectly absorbing blackbody.

Using Eq.\:\ref{d} and $p_0$, we can determine the moment of inertia $I_{\perp}$.  We can then infer the size of the particle from
\begin{equation}
\label{eq:momentinertia}
I_{\perp}=\int d a\, r^{2} \rho= m \frac{\int d a\, r^{2} \rho}{\int d a\, \rho}=m \left\langle r_{m}^{2} \right\rangle,
\end{equation}
where $\rho$ is the local two-dimensional mass density of the flake, $\left\langle r_{m}^{2} \right\rangle$ is the mean square size of the flake weighted by its mass density, and the integrals are over the surface of the flake.

\subsection{Reorientation of the flake from  gas friction} \label{subsec:reorientation}
Because $\bm{\hat{z}} \| \bm{E}_{\mathrm{rf}}$, $N_{z}=0$.  This simple fact has important consequences for flake orientation after locking, since torques from the applied electric field cannot counteract torques along $\bm{\hat{z}}$ arising from friction with residual gas in the vacuum chamber, and consequently the flake is driven to a state in which $L_{z}=0$.  To model this effect we assume: 
\[\bm{N}_{g}=-\bm{L}/\tau_{g}=-\left|L \right|\bm{\hat{n}}/\tau_{g},\] 
where $\tau_{g}$ is a phenomenological relaxation time.  If the flake is to remain locked, however, this torque must be partially counteracted by a torque from the applied rf electric field:
\[-\left|L \right|/\tau_{g}+N_{n}=0.\]
But, using Eqs.\:\ref{a} and \ref{b}:
\[N_{\theta}=N_{n} \cot \theta.\]
Because the flake is locked, only the orientation of $\bm{L}$, but not its magnitude, can change.  
Thus:
\[N_{\theta}=\dot{L}_{\theta}=\left|L \right| \dot{\theta} =\frac{1}{\tau_{g}}\cot \theta \left|L \right|.  \]
This equation has the solution:
\begin{equation}\label{h}
\theta (t) =\cos^{-1}  ( e^{-t/\tau_{g}} ).  
\end{equation}
In the presence of background gas, the locked flake reorients towards $\theta=90^{\circ} $, where 
$\bm{\omega_{n}}$ lies in the x-y plane.
\begin{figure}
\begin{center}
\includegraphics[scale=1,draft=false]{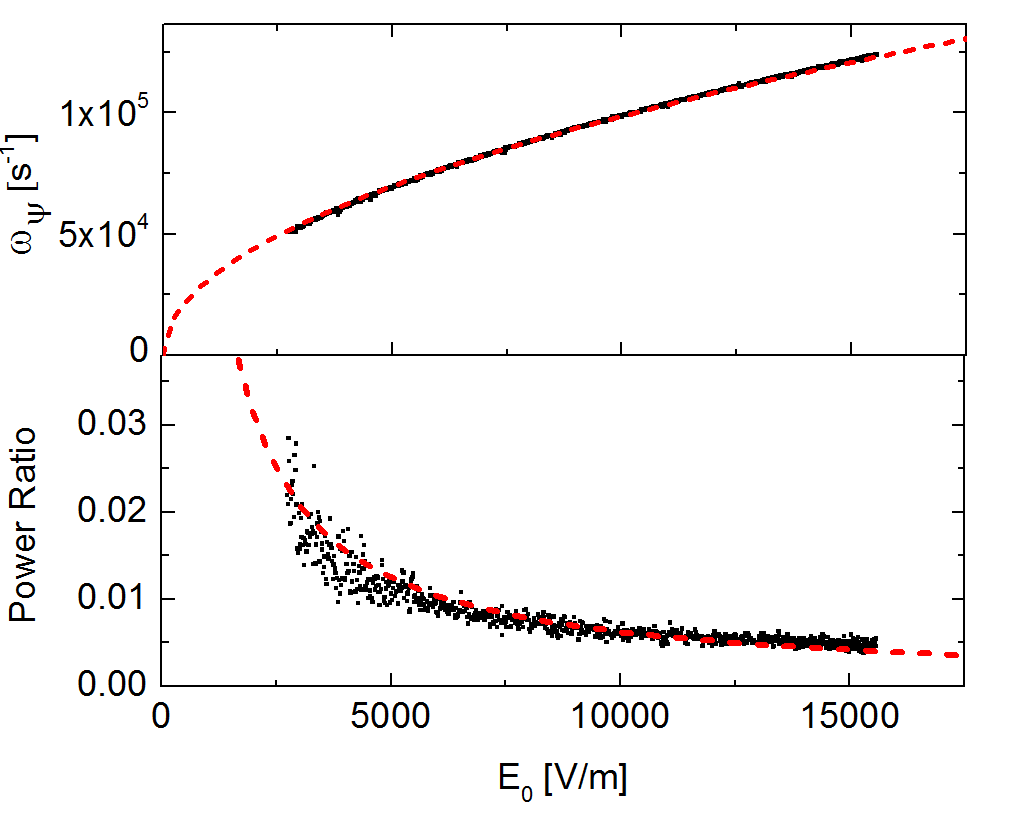}
\end{center}
\caption{(a) Sideband separation $\omega_{\psi}$ as a function of rf electric field, measured after the flake is locked and oriented at $\theta \cong 90^{\circ}$.  The rf electric field strength is varied over the course of 3 hours, so that the particle remains stably locked.  The square root dependence shows that $E_0$ is coupling to a permanent electric dipole on the flake.  (b) Sideband power ratio as a function of $E_0$ 
in the same conditions.  From this data, we can calculate the electric dipole moment, $p_{0}=2.6 \times 10^{-22} ~\mathrm{C} \, \mathrm{m}$ (Eq.\:\ref{eq:sidebandpower}) and the flake moment of inertia, $I_{\perp}=1.4 \times 10^{-28} ~\mathrm{kg} \, \mathrm{m}^{2}$ (Eq.\:\ref{d}).}
\label{fig:sidebandgap}
\end{figure}

\subsection{Flake precession of $\bm{\omega_{n}}$ around $\bm{\hat{z}}$} \label{subsec:precession}
Eq.\:\ref{b} implies that when $\theta\neq 90^{\circ} $, $N_{\phi}\neq 0$ and consequently the axis of rotation of the flake must precess around $\bm{\hat{z}}$. We assume that this precession is slow compared to the 
$\omega_{\psi}$ oscillations but rapid compared to the decay in time of $\theta$ due to friction.  Thus we use Eq.\:\ref{b} with a constant value for $\theta$ and an averaged value for $ \cos \psi$:
\[N_{\phi}=+\frac{1}{2}p_{0}E_{0}\cos \theta \langle\cos \psi \rangle =\dot{L}_{\phi}=I_{\perp} \omega_{n}  \dot{\phi} \sin \theta, \]
from which we get:
\begin{equation}\label{i}
\omega_{\phi} \equiv \dot{\phi}=\frac{\Omega^{2}}{\omega_{n}} \langle\cos \psi \rangle \cot \theta.
\end{equation}
Precession ceases when $\theta \to 90^{\circ}$ or when the particle loses lock and $\langle\cos \psi \rangle \to 0$ .  When thermal fluctuations of $\psi$ are negligible, $\langle\cos \psi \rangle=1$.
\begin{figure}
\begin{center}
\includegraphics[scale=0.26,draft=false]{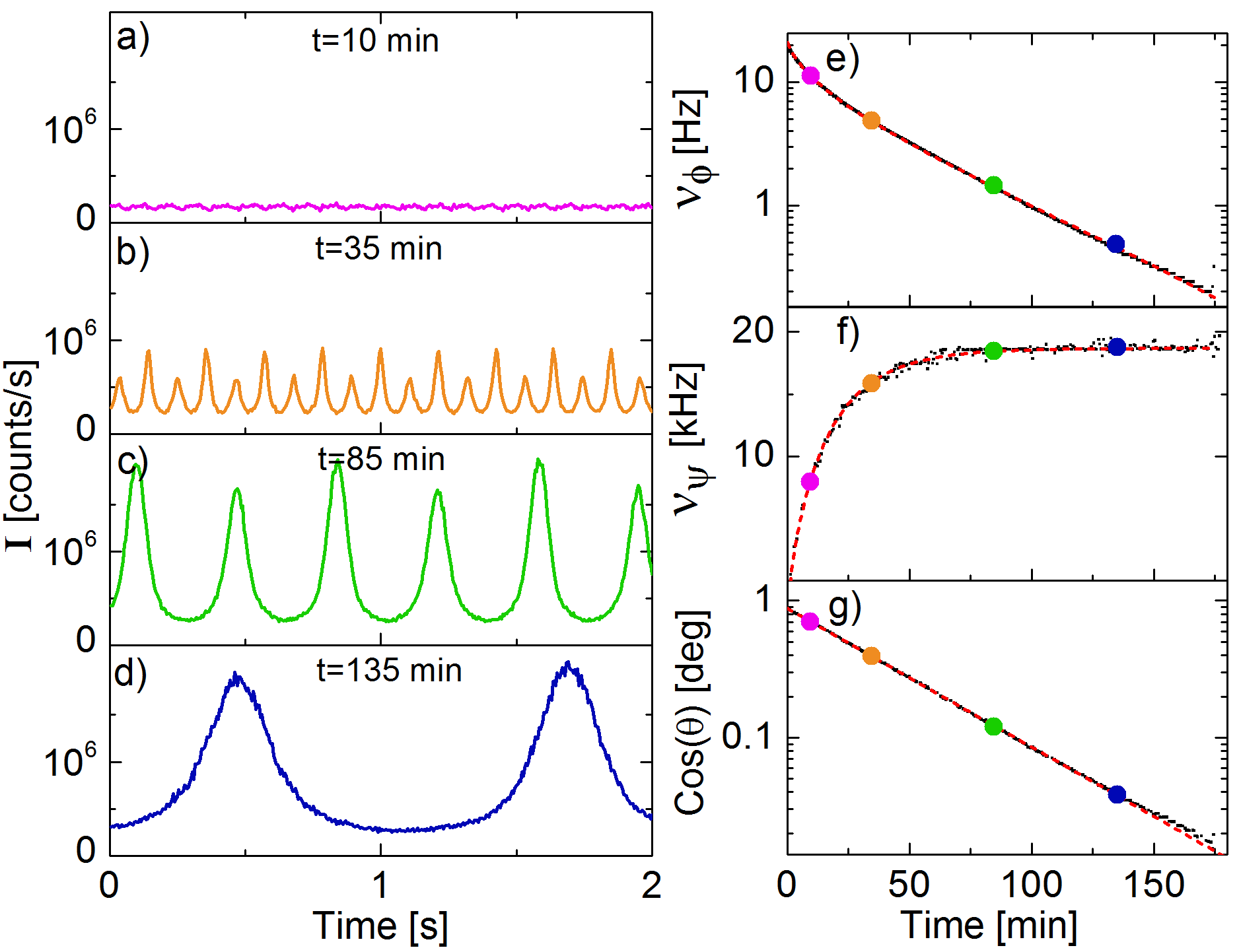} 
\end{center}
\caption{(a-d) Time evolution of low frequency light scattering from the flake after locking at time t=0. Fluctuations in the signal increase in magnitude and become regularly spaced peaks whose period increases with time.  Plot colors correspond to colored dots on plots (e-g).  (e) Precession frequency ($\nu_{\phi}=\omega_{\phi}/(2\pi)$) determined from the lowest Fourier component of the optical data.  (f) Sideband frequency ($\nu_{\psi}=\omega_{\psi}/(2\pi)$) measured as a function of time after locking.  (g) Orientation of the flake, inferred from measurements of 
$\omega_{\psi}$ and 
$\omega_{\phi} $ and calculated using Eqs.\:\ref{d} and \ref{i}. Red dashed line is behavior expected from Eq.\:\ref{h}. Note log scales in (e) and (g). The data in this figure is from a different sample than the data in Fig.\:\ref{fig:dimensions}; although the peak heights and periods are different, the dynamics are qualitatively similar.}
\label{fig:angledynamics}
\end{figure}

\section{Comparing the model to observations} \label{sec:observations}

We have observed stable locking behavior in 20 individual particles.  Data in Figs.\:\ref{fig:sidebandgap}, \ref{fig:dimensions}, \ref{fig:orientationdata}b, and \ref{fig:orientationtheory}, as well as in Table\:\ref{tab:calculated}, are from a single particle (labeled Sample D) on which we collected the most complete set of data, while data in Figs.\:\ref{fig:locking}, \ref{fig:angledynamics}, and \ref{fig:orientationdata}a are from three other particles (labeled Samples A, B, and C). 

Fig.\:\ref{fig:sidebandgap} shows data on a particle taken several hours after locking, when it has completely stabilized.  The sideband spacing (Fig.\:\ref{fig:sidebandgap}a) clearly has the square root dependence on $E_0$ predicted from Eq.\:\ref{d}, and it is consequently possible to estimate $p_{0}$ from the data.  Also, the power in the sidebands relative to the harmonics (Fig.\:\ref{fig:sidebandgap}b) shows the behavior expected from the model (Eq.\:\ref{eq:sidebandpower}), although thermal fluctuations are likely significant.  From this data, we can estimate the moment of inertia $I_{\perp}$ from Eq.\:\ref{eq:sidebandpower}.  

From Eq.\:\ref{eq:momentinertia}, we can estimate the size of the flake. The mass of the particle can
be estimated from the Brownian motion of particle in the trap at relatively high pressure (30 mTorr), assuming particle motion is thermalized and T $\sim$ 300 K [8]. For this sample, we find $m=3.7\times10^{-17}$ kg, which gives $\left\langle r_{m}^{2} \right\rangle = 3.7 \times 10^{-12}~\mathrm{m}^2$. The actual shape of our platelet is unknown and may be quite irregular.  However, if we make the assumption that it is a circular disc of uniform thickness, then its area is $2\pi \left\langle r_{m}^{2} \right\rangle = 2.3 \times 10^{-11}~\mathrm{m}^2$.  Taking the areal density of single-layer graphene to be $7.4 \times 10^{-7} \mathrm{kg}\,\mathrm{m}^{-2}$, this area is consistent with a particle of $\sim$2 layers.  

The low frequency behavior of the scattered light intensity is shown in Fig.\: \ref{fig:angledynamics}(a-d).  After the flake is locked and the circularly polarized laser is turned off, regular oscillations appear whose intensity increases and frequency decreases with time. As the predominate frequency of these oscillations approaches $\sim$1 Hz, the intensity of the signal stabilizes and appears as two peaks with slightly different amplitudes (Fig.\:\ref{fig:angledynamics}(b-d)).  We interpret this behavior as a consequence of slow precession.  The evolution of this precession frequency over time, determined from the frequency of the lowest Fourier component of the intensity data, is plotted in Fig.\:\ref{fig:angledynamics}e.  Also plotted (Fig.\:\ref{fig:angledynamics}f) is the data taken simultaneously for the sideband spacing.  Using these experimental values of $\omega_{\psi}$ and 
$\omega_{\phi}$, and using Eqs.\:\ref{d} and \ref{i}, we can determine $\cos \theta$ (see Fig.\:\ref{fig:angledynamics}g).  The fit to the expected behavior in the model (Eq.\:\ref{h}) is excellent.

\section{Estimating nanoplatelet dimensions using optical signal from low frequency precession} \label{sec:opticaldimensions}

We can obtain an alternate estimate of the flake's dimensions from the shape of the optical signal from low frequency precession seen in Fig.\:\ref{fig:angledynamics}.  Crudely, the double peak behavior observed when $\theta \to 90^{\circ}$ is a consequence of the flake behaving as a slowly precessing mirror that reflects the light from the source to the detector twice each full precession cycle. The intensity peaks occur when the flakes are oriented parallel to the mirroring plane (Fig.\:\ref{fig:apparatus}b).  The width of the peaks is finite because the size of the flake is comparable to the wavelength of the light.  

\begin{figure}
\begin{center}
\includegraphics[scale=1,draft=false]{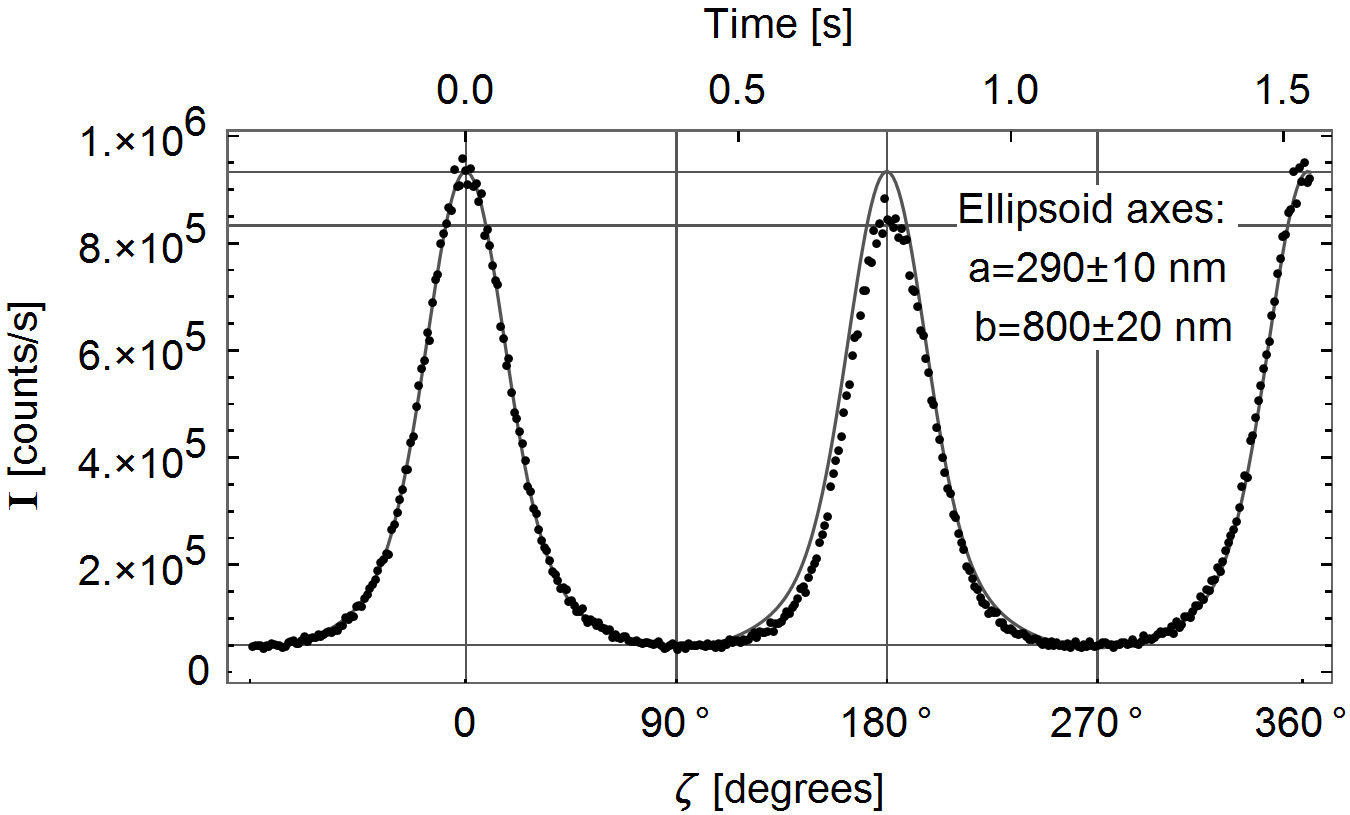} 
\end{center}
\caption{Scattered light intensity data for one period of precession. $\zeta=0$ corresponds to the orientation where the flake is in the mirroring plane and the signal is maximized.  Data for one half-period $(-90^{\circ} < \zeta < 90^{\circ}$) is fit to Eqs.\:\ref{j} and \ref{eq:besselellipse} in order to approximate the dimensions of the flake. The best-fit curve and derived values of the semiminor and semimajor axes $a$ and $b$ are shown.  The fit is excellent over the first half-period but shows the small asymmetry in the second peak.}
\label{fig:dimensions}
\end{figure}

Low frequency precession occurs around the direction of $\bm{E}_{\mathrm{rf}}$, assumed to be along the z axis.  Assuming perfect alignment of the trap electrodes, $\bm{ \tilde{k}} \perp \bm{\hat{z} }$, and thus 
$\tilde{k}_{z}=0 $.  As a consequence of time reversal symmetry, 
$F (\bm{\tilde{k}})=F (-\bm{\tilde{k}})  $, so $180^{\circ} $ rotations of the flake around the
z-axis leave the form factor unchanged.  This symmetry is not exact in the data shown in Fig.\:\ref{fig:dimensions} (the intensity peaks are not of equal amplitudes), probably because of small deviations from orthogonality of $\bm{\tilde{k}}$  and $\bm{E}_{\mathrm{rf}}$. 
  
For low frequency data, the rapid time dependence of $I(t)$ due to rotation around $\bm{\hat{n} } $ is averaged out, and the observed intensity only depends on the angle $\zeta$ between
$\bm{\hat{n} } $ and $\bm{\tilde{k}} $. 
While the nanoplatelet is likely to be irregularly shaped, measurement of the angular dependence of scattering provides an estimate of its size\citep{Gordon2007}\cite{Schiffer1979}\citep{LeVine1983}. 
If we assume that the nanoplatelet is a circular disk with diameter $2r_{0}$:  
\begin{equation}\label{j}
F(k_{\|},r_{0})=(\pi r_{0}^{2})^{2} \left\lbrace \frac{2 J_{1}(k_{\|}r_{0})}{k_{\|}r_{0}} \right\rbrace^{2},  
\end{equation}
where $J_{1} $ is the first order Bessel function, and $k_{\|} = |\bm{\tilde{k}}_{\|}| $ is the  component of $\bm{\tilde{k}}  $ in the plane of the disk:
\[k_{\|}=\frac{4 \pi}{\lambda_{0}} \sin \frac{\gamma}{2} \sin \zeta. \]
$ \lambda_{0} = 532$ nm is the wavelength of the incident light and $\gamma $ 
is the angle through which the light scatters, which is $38^{\circ}$ in the experiments (Fig. 1).

We can approximate the scattering from nonuniform objects by considering the form factor for elliptical nanoplatelets, which can be readily obtained from the result for circular disks via a coordinate transformation.  For an ellipse with semiminor and semimajor axes $a$ and $b$, the following substitutions into Eq.\:\ref{j} yield the appropriate values:
\[r_{0}^{2} \to ab\]
and:
\begin{equation}
\label{eq:besselellipse}
k_{\|}r_{0} \to k_{\|} \sqrt{\frac{1}{2}(a^{2}+b^{2})+\frac{1}{2}(a^{2}-b^{2}) \cos 2 \Phi},
\end{equation}
where $\Phi $ is the angle between $\bm{\tilde{k}}_{\|}$  and the major axis of the ellipse.  Because of the rapid rotation, averaging over $\Phi $ yields the appropriate result for interpreting the low frequency optical signal.

Using this elliptical approximation, we can fit the experimental peak amplitudes to determine dimensions of the flake. We perform a fit of the function in Eq.\:\ref{j}, with the substitution for $r_0$ as shown in Eq.\:\ref{eq:besselellipse}, in which the function is averaged over one full in-plane rotation of the particle (\textit{i.e.} $0<\Phi<2\pi$) during each iteration of the fitting algorithm.  Due to the small asymmetry of the peaks, we fit only half the period of precession.  Fig.\:\ref{fig:dimensions} shows the data and fit curve, from which we estimate $a=290$ nm and $b=790$ nm for the semiminor and semimajor axes respectively.

These dimensions give a flake area of approximately $\pi ab = 7.2 \times 10^{-13}~\mathrm{m}^2$, which implies a layer number of $\sim$70.  This is a significantly smaller area than that obtained using the measured moment of inertia in Sec.\:\ref{sec:observations}, and the likely uncertainty of $\sim$30\% in the mass does not account for the discrepancy.  We have thus far assumed that only the size of the flake contributes to the width of the scattered light peaks; however, there are two likely sources of systematic error that could cause added broadening of the peaks, leading to underestimation of the areal dimensions of the platelet.  First, the above calculation assumes that the acceptance angle of the lens is infinitely small.  We estimate that accounting for the acceptance angle will increase the semimajor axis $b$ by a factor of $\sim$1.3.  More importantly, though, in the above calculations, any high frequency precession of the instantaneous axis of rotation was neglected, as were flexural modes of the membrane. Rapid wobbling or bending of the membrane in this manner would broaden the peaks and lead to an underestimation of its dimensions in the previous calculation.

\begin{table}
\begin{center}
\setlength{\tabcolsep}{5pt}
\renewcommand{\arraystretch}{1.5}
\begin{tabular}{ccc}
\hline 
Quantity & Value &  \\
\hline \hline 

$q/m$ & 6.1 & $\mathrm{C}\,\mathrm{kg}^{-1}$  \\

$m$  & $3.7 \times 10^{-17}$ & kg  \\ 

$\Omega$  & $1.27 \times 10^{5}$ & $\persec$  \\

$p_{0}$ & $2.6 \times 10^{-22}$ & $\mathrm{C}\,\mathrm{m}$  \\

$I_{\perp}$ & $1.4 \times 10^{-28}$ & $\mathrm{kg} \, \mathrm{m}^2$ \\

\hline

$\alpha_{dia}$ & $-7.6 \times 10^{-17}$ & $\mathrm{J} \, \mathrm{T}^{-2}$ \\

$\alpha_{rot}$ & $2.9 \times 10^{-27}$ & $ \mathrm{J} \, \mathrm{s} \, \mathrm{T}^{-1}$ \\

$g$ & 7.6 & \\
\hline

Area & $2.3 \times 10^{-11}$ & $\mathrm{m}^2$ \\

Number of layers & $\sim$2 & \\
\hline
\hline
\end{tabular}
\end{center}
\caption{Calculated quantities for Sample D. \textit{Top section:}  $q/m$ is derived from Eq.\:\ref{eq:eigenf}. The measurement of $m$ is described in Sec.\:\ref{sec:observations}. $\Omega$ is derived from Eq.\:\ref{d}, measured after particle has fully reoriented and $\sin{\theta} \approx 1$.  $p_0$ comes from fitting data in Fig.\:\ref{fig:sidebandgap}b to Eq.\:\ref{eq:sidebandpower}.  $I_{\perp}$ comes from fitting data in Fig.\:\ref{fig:sidebandgap}a to Eq.\:\ref{d}. \textit{Middle section:}  $\alpha_{dia}$ is derived from Eq.\:\ref{l}. $\alpha_{rot}$ is derived from Eq.\:\ref{m}. $g$ is derived from Eq.\:\ref{eq:gfactor}. \textit{Bottom section:} The flake area and layer number are estimated in Sec.\:\ref{sec:observations}.  For the last two quantities, we assume that the flake is circular and uniform in thickness; however, the actual shape is unknown. }
\label{tab:calculated}
\end{table}

\section{Interactions with a magnetic field when $\theta \to 90^{\circ}$} \label{sec:magnetic}
\begin{figure}
\begin{center}
\includegraphics[scale=1,draft=false]{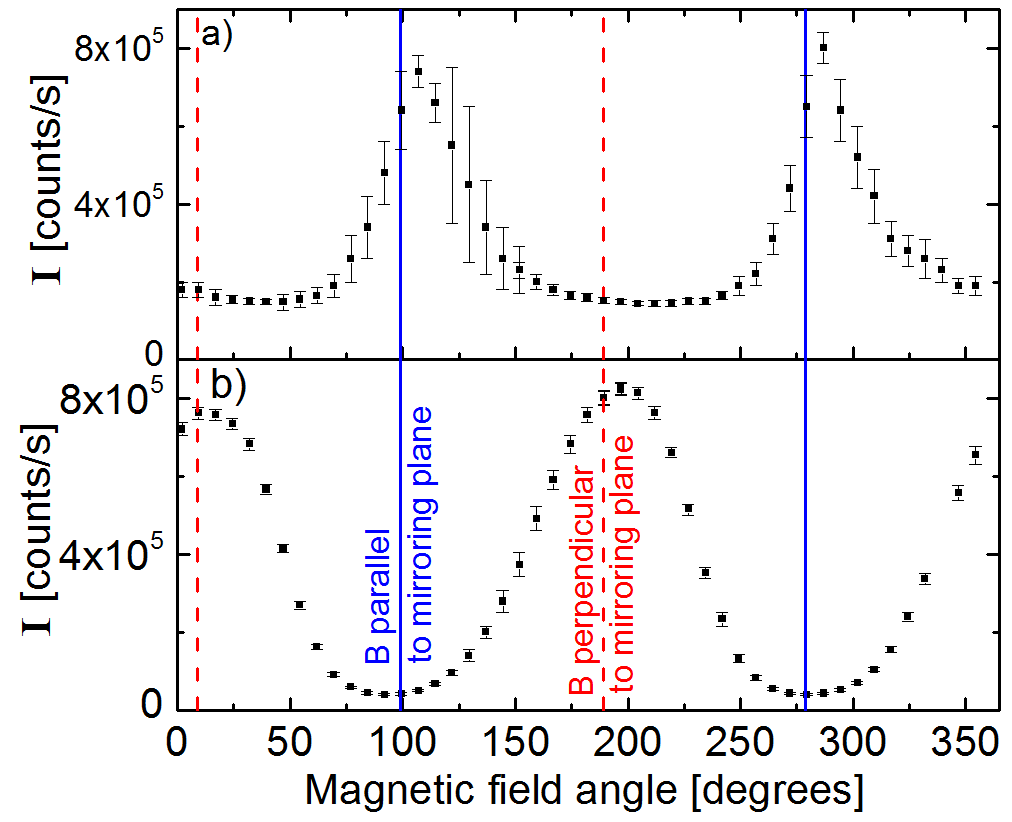} 
\end{center}
\caption{Control of trapped flake orientation by rotating the direction of the applied magnetic field.  Field magnitude at the trap center varies from 4.5-7.2 mT as the dipole magnets are rotated.  The zero of the magnetic field angle is defined such that the magnets are pointing collinearly (as shown in Fig.\:\ref{fig:apparatus}b).  As the magnets are rotated clockwise in tandem, the magnetic field rotates \textit{counter-}clockwise.  Data points represent mean intensity.  Error bars represent the approximate intensity variation arising from oscillation of the particle about a fixed orientation.  (a) Data from Sample C, with drive frequency $\nu_0=\nu_n=\omega_n/ \left(2\pi\right)=20$ MHz.  Orientations with a maximum intensity occur when B is parallel to the mirroring plane (solid vertical lines). (b) Data from Sample D, with $\nu_0=40$ MHz.  Maximum intensity occurs when B is perpendicular to the mirroring plane (dashed vertical lines). Data in plots (a) and (b) is taken from different samples; however, we have observed the transition from parallel to perpendicular alignment in a single particle.}
\label{fig:orientationdata}
\end{figure}

We have observed that when $\theta \to 90^{\circ}$ flake precession inevitably ceases. The flake becomes orientationally trapped and subsequently oscillates around a fixed orientation.  Using rotatable permanent dipole magnets located outside the vacuum chamber (Fig.\:\ref{fig:apparatus}b), we have established that the direction of $\bm{B}$ in the xy-plane determines the trapping direction.  In Fig.\:\ref{fig:orientationdata}, the average intensity of light scattered from a trapped flake is plotted as a function of the direction of $\bm{B}$ in the xy-plane.  Two maxima occur during a full $360^{\circ}$ rotation of the magnets, indicating that the flake has also rotated $360^{\circ}$.

We see the flake assume two distinct orientations with respect to the direction of $\bm{B}$, depending on its rotation frequency.  While the flake is locked, the rotation frequency $\nu_n=\omega_n/\left(2\pi\right)$ is equal to the rf drive frequency $\nu_0=\omega_0/\left(2\pi\right)$ (over times longer than the $\sim$0.1 ms period of the $\omega_\psi$ oscillations), so we can directly vary $\nu_n$ by varying $\nu_0$.  

At relatively low $\nu_n$ (Fig.\:\ref{fig:orientationdata}a), two intensity maxima occur when $\bm{B}$ is in the mirroring plane (Fig.\:\ref{fig:apparatus}b), implying that the flake is aligned such that $\bm{B}$ lies in the flake plane.  This behavior is consistent with a diamagnetic response of a 2D flake\citep{Ominato2013}\cite{Uyeda2001}.  

At higher $\nu_n$ (Fig.\:\ref{fig:orientationdata}b), however, the two intensity maxima appear when $\bm{B}$ is \textit{normal} to the mirroring plane, implying that the flake is aligning normal to $\bm{B}$.  This behavior is to be expected if the flake's magnetic moment arises not from diamagnetism but from the charged nanoplatelet's rapid rotation.  

In order to understand these observations, we construct a model to account for the two distinct responses of the flake to the magnetic field.  First, we consider the diamagnetic response of a 2D flake.  The flake can experience a torque from a magnetic field if it has an anisotropic magnetic polarizability. For a thin flake, we assume that the induced dipole is perpendicular to the flake plane and proportional to the component of $\bm{B}$ perpendicular to the plane:
\[\bm{\mu}_{dia}=\alpha_{dia} \bm{\hat{n}} \left(  \bm{\hat{n}} \cdot \bm{B} \right).  \]
Note that if $\alpha_{dia}<0$, the induced dipole points away from $\bm{B}$ as would be expected for diamagnetic materials\cite{Uyeda2001}\citep{Sepioni2010}.

Second, we consider the magnetic moment $\bm{\mu}_{rot}$ created by the charged nanoplatelet's rapid rotation. If the magnetic moment arises from the current created by the charge fixed on the spinning particle, then:
\[\bm{\mu}_{rot}= \alpha_{rot}\bm{\omega}=\alpha_{rot} \omega_{n} \bm{\hat{n}}. \]
Assuming that $\theta \cong 90^{\circ}$ and $\bm{B} = B_0 \bm{\hat{x}}$: 
\[
\bm{N} =\bm{\mu}  \times \bm{B}=B_{0}
\begin{vmatrix}
\bm{\hat{\theta}} & \bm{\hat{\phi}} & \bm{\hat{n}}  \\ 
0 & 0 &\alpha_{dia} B_{0} \cos \phi +\alpha_{rot} \omega_{n}  \\
0 & -\sin \phi & \cos \phi 
\end{vmatrix},
\]
with $\bm{\mu}=\bm{\mu}_{dia}+\bm{\mu}_{rot}$.  Thus:
\[N_{\theta}=-I_{\perp} \omega_{n} \dot{\theta}'=\alpha_{dia} B_{0}^{2} \sin \phi \cos \phi+ 
\alpha_{rot}B_{0} \omega_{n}\sin \phi,\]
where $\theta'=90^{\circ}-\theta$.   For small $\theta'$ we use Eq.\:\ref{i} 
with $\langle\cos \psi \rangle\cong 1$ to get:
\[\dot{\phi}=\frac{\Omega^{2}}{\omega_{n}} \theta'.\]
Consequently:
\begin{equation}\label{k}
\ddot{\phi}=-\frac{\Omega^{2}}{I_{\perp} \omega_{n}^{2}} \left\lbrace B_{0}^{2} \alpha_{dia}\sin \phi \cos \phi+B_{0}\omega_{n}\alpha_{rot} \sin \phi \right\rbrace.  
\end{equation}
When the second term in Eq.\:\ref{k} is neglected, the locations of stable points of oscillation depend on the sign of $\alpha_{dia}$.  When $\alpha_{dia}<0$, they are at $\phi= \pm 90^{\circ}$, and for small amplitudes the oscillations have angular velocity: 
\begin{equation}\label{l}
\omega_{dia}=\sqrt{\frac{\Omega^{2}B_{0}^{2}\left|\alpha_{dia} \right| }{I_{\perp} \omega_{n}^{2}}}.
\end{equation}
When the first term is neglected, stable oscillations occur near $\phi=0$.  For small $\phi$, the oscillations have  angular velocity:
\begin{equation}\label{m}
\omega_{rot}=\sqrt{\frac{\Omega^{2} B_{0}\alpha_{rot}}{I_{\perp} \omega_{n}}}.
\end{equation}
The expected oscillation frequencies in Eqs.\:\ref{l} and \ref{m} have different functional dependences on
$B_{0}$ and $\omega_{n}$ which allow us to distinguish the mechanisms with data\footnote{Another mechanism that can introduce torque and trap the orientation of the flake comes from the rotation of the Earth.  For this mechanism trapping oscillation angular velocities will be
\unexpanded{$\approx\Omega\sqrt{2\pi/\omega_{n}D}$}, where $D$ is the period of the Earth's rotation.  For our experiments, this would lead to periods of order 100 sec., much slower than what arises from magnetic fields.}.

\begin{figure}
\begin{center}
\includegraphics[scale=1,draft=false]{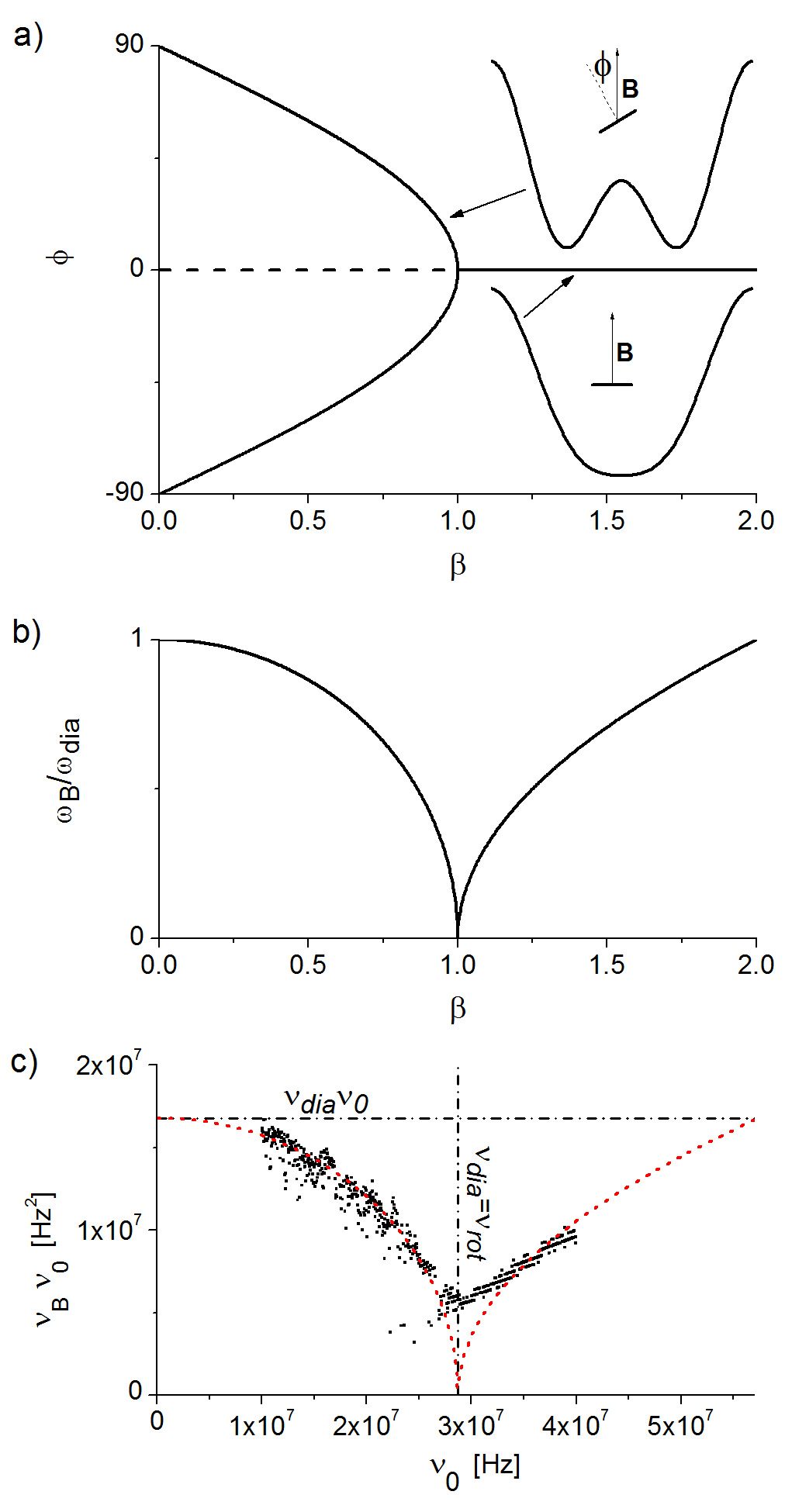} 
\end{center}
\caption{Magnetic orientation of a spinning flake. (a) shows predicted orientation angles $\phi$ of flake in diamagnetism-dominant $\left( \beta < 1 \right)$ and rotation-dominant $\left( \beta > 1 \right)$ regimes. (b) shows the dependence of the quotient $\omega_B / \omega_{dia}$ (see Eqs.\,\ref{eq:oscfreq1} and \ref{eq:oscfreq2}) in each regime. (c) shows experimental data, in which the rf drive frequency $\nu_0 = \nu_n = \omega_{n}/\left(2\pi\right)$ is varied (at a rate of 400 $\mathrm{Hz}\,\persec$) and the oscillation frequency $\nu_B=\omega_B/\left(2\pi\right)$ is measured.  Best-fit curves (red dashed lines) corresponding to Eqs.\:\ref{eq:oscfreq1} and \ref{eq:oscfreq2} are overlaid.  The dash-dotted horizontal line marks the constant value of the product $\nu_{dia} \nu_0 = 1.68 \times 10^7 ~\mathrm{Hz}^2$.  The dash-dotted vertical line marks the drive frequency $\nu_0=2.87 \times 10^7 ~\mathrm{Hz}$ at which the contributions to the oscillation frequency from diamagnetism and rotation are equal: $\nu_{rot}=\nu_{dia}=0.58$ Hz.  For this data, the external magnets are fixed at an angle of $+28^{\circ}$ from the zero defined in Fig.\:\ref{fig:orientationdata}.  The resulting field strength near the particle is approximately B=7 mT.}
\label{fig:orientationtheory}
\end{figure}

When both terms are comparable, we need to solve:
\begin{equation}
\label{n}
\frac{\ddot{\phi}}{\omega_{dia}^{2}}=f\left(\phi\right)=\sin \phi \cos \phi -\beta \sin \phi,
\end{equation}
where $\beta \equiv \omega_{rot}^{2}/\omega_{dia}^{2}$.  Solutions are stationary when the right hand side is zero:
\[\sin \phi = 0 \mathrm{~~~or~~~} \cos \phi =\beta.\]
To obtain the frequency of small oscillations, we determine the first derivative of $f(\phi)$ at the stationary angles:
\[f' \left( \sin \theta = 0 \right)= 1\pm \beta ~~~~~\mathrm{and:}~~~~~ f' \left(\cos \phi= \beta \right)=\beta^{2}-1. \]
Thus, the angular velocities for small oscillations, $\omega_{B}$, are
\begin{align}
\label{eq:oscfreq1}
\omega_{B}=\omega_{dia} \sqrt{1-\beta^{2}}~~~~~~\left(\beta < 1 \right)
\end{align}
and
\begin{align}
\label{eq:oscfreq2}
\omega_{B}=\omega_{dia} \sqrt{\beta-1}~~~~~~\left(\beta \geq 1 \right).
\end{align}
These solutions are plotted in Figure \ref{fig:orientationtheory}(a,b). Since the confining potential is not simple harmonic, deviation from Eqs.\:\ref{eq:oscfreq1} and \ref{eq:oscfreq2} will be substantial, especially near $\beta=1$.

\section{Estimates of the nanoplatelet magnetic properties} \label{sec:magdipole}

We test the above model by varying the rf drive frequency $\nu_0$ (slowly enough so that the particle remains locked and $\nu_0=\nu_n$) and observing the oscillation frequency $\nu_B=\omega_B/\left(2\pi\right)$. The external magnetic field is held constant at $B_0=7$ mT. In Fig.\:\ref{fig:orientationtheory}c, we plot the product $\nu_B \nu_0$.  In the range $10 ~\mathrm{MHz} < \nu_0 < 25 ~\mathrm{MHz}$, the data can be fit fairly well by Eq.\:\ref{eq:oscfreq1}.  

\subsection{Diamagnetic polarizability} \label{subsec:diamagneticdata}

We first examine the data in the regime where the diamagnetic contribution is dominant, where $\nu_0 \to 0$. The fit in Fig.\:\ref{fig:orientationtheory}c suggests that in this limit, $\nu_B \nu_0 \to 1.68 \times 10^7 ~ \mathrm{Hz}^2$.  In the same limit, Eq.\:\ref{eq:oscfreq1} becomes $\omega_B=\omega_{dia}$, and we can use Eq.\:\ref{l} to obtain $\alpha_{dia}=-7.6 \times 10^{-17} ~ \mathrm{J}\,\mathrm{T}^{-2}$.

We can make a direct comparison of our data with the ``magnetic shield" model for graphene diamagnetism developed by Koshino \textit{et.\,al.} \citep{Koshino2009}. In this theory, diamagnetic currents attenuate a magnetic field perpendicular to the flake by a constant factor $1-f,$ where:
\begin{equation}
f=\frac{2\pi g_{v} g_s e^2 \nu_f}{16 \hbar c^2}.
\end{equation}
$g_{v}$ and $g_s$ are the valley and spin degeneracies respectively and $\nu_f$ is the constant Fermi velocity for graphene.  For $g_{v}=g_s=2$ and $\nu_f/c \approx 10^{-3},$ $f=4 \times 10^{-5}$. For a circular disk of such a material\citep{Arvas},
\[\alpha_{dia}=-\frac{8}{3\mu_0}r^3f,\]
where $r$ is the disc radius and $\mu_0=4 \pi \times 10^7 ~\mathrm{H}\,\mathrm{m}^{-1}$.  

Most of our calculations have been made without assumptions as to the shape of the particle; however, if we assume that it is a circular disc, then using the mean square size of the platelet calculated from the moment of inertia in Sec.\:\ref{sec:observations}, with $r^3 \rightarrow \left(2\left\langle r_m^2 \right\rangle \right)^{3/2}$, Koshino's formula gives  $\alpha_{dia}= -1.7 \times 10^{-15} \mathrm{J}\,\mathrm{T}^{-2}$, about 20 times larger in magnitude than the measured value listed in Table \ref{tab:calculated}. Since we have used the larger of the two size estimates in Secs.\:\ref{sec:observations} and \ref{sec:opticaldimensions}, this may be an overestimate.  Improved measurements of the size and mass of the flake will be necessary for more accurate comparisons of data and theory.  It should also be noted that Koshino's theory assumes that the material is at the Dirac point, which may be untrue for our platelet because it has a net charge.

\subsection{Rotationally induced magnetic moment} \label{subsec:rotationalmagdata}

Returning to the data in Fig.\:\ref{fig:orientationtheory}c, we see that the minimum in the product $\nu_B \nu_0$ occurs around a drive frequency of $\nu_{min}=2.87 \times 10^7 ~\mathrm{Hz}$.  We interpret this as the turning point between the diamagnetic and rotational regimes, where the contributions $\omega_{dia}$ and $\omega_{rot}$ are equal.  Combining Eqs.\:\ref{l} and \ref{m} gives

\[ \frac{\omega_{dia}^2}{\omega_{rot}^2}=\frac{B_0 \left| \alpha_{dia} \right|}{\omega_{min} \alpha_{rot}}=1, \]
from which we obtain $\alpha_{rot}=2.9 \times 10^{-27} ~ \mathrm{J}\,\mathrm{s}\,\perTesla$.

Assuming that the magnetic dipole moment arises from rotation of the charge on the rotating nanoplatelet, the contribution from rotation, $\bm{\mu}_{rot}$, can be obtained from\citep{Jackson3rd}: 
\[\bm{\mu}_{rot}=\frac{\bm{\omega_{n}}}{2} \int d a\, r^{2} \sigma=\frac{q \bm{\omega_{n}}\int d a\, r^{2}\sigma}{2 \int d a\,  \sigma}=\frac{q\omega_{n}}{2}\left\langle r_{q}^{2} \right\rangle  \bm{\hat{n}}.\]
Here, the integrals are over the surface of the flake, $r$  is the distance from the axis of rotation, and $\sigma$ is the local charge density on the flake surface; $q$ is the total charge on the flake and $\left\langle r_{q}^{2} \right\rangle $ is the mean square size of the flake weighted by the charge density. Thus: $\alpha_{rot}=q \left\langle r_{q}^{2} \right\rangle /2$ and:
\[\omega_{rot}=\sqrt{\frac{\Omega^{2} B_{0} q \left\langle r_{q}^{2} \right\rangle}{2 I_{\perp} \omega_{n}}}.\]
If we recall the definition of the moment of inertia in Eq.\:\ref{eq:momentinertia} and define  
$g \equiv \left\langle r_{q}^{2} \right\rangle / \left\langle r_{m}^{2} \right\rangle$  
then we can obtain an expression for $\omega_{rot}$:
\begin{equation}
\omega_{rot}=\Omega \sqrt{ \frac{ g }{2 \omega_{n}} \frac{q B_0}{m}}.
\label{eq:gfactor}
\end{equation}
For the data shown in Fig.\:\ref{fig:orientationtheory}c, $B_0=7$ mT, and thus $q B_0/m=0.04 ~\persec$; also, at the turning point, $\omega_{rot}=3.68 ~\persec$.  Using Eq.\:\ref{eq:gfactor}, we obtain $g=7.6$.

The g-factor that we have inferred from our data is substantially in excess of that which would be expected for a spinning charged conducting plate, where $g \approx 1-2$ \footnote{For a spinning conducting circular disk, \unexpanded{$g=\left\langle r_q^2 \right\rangle / \left\langle r_m^2 \right\rangle = 4/3$}.  If \textit{all} charge is located at the perimeter of a circular disc of uniform mass density, then \unexpanded{$\left\langle r_q^2 \right\rangle / \left\langle r_m^2 \right\rangle =2$}.}.  It is thus possible that we are observing the Barnett effect, a magnetic moment induced by the rotation that is present even when there is no fixed surface charge, \textit{i.e.} $q \to 0$ \citep{Barnett1935}\citep{Chudo2014}\citep{Ono2015}\citep{Baasanjav2015}.  For electrons of mass $m_e$ and charge $e$, the effective magnetic field induced by rotation is 
\[ B_{rot}=\frac{m_e}{e}\omega_n, \] 
which for $\nu_0=\omega_n/ \left(2\pi\right) = 30$ MHz is $B_{rot}=1.1 \times 10^{-3}$ T. This field is quite comparable to that used in the experiments and is suggestive that the Barnett effect may be important.

The measurements of $\alpha_{dia}$ and $\alpha_{rot}$ listed in Table \ref{tab:calculated} imply (for B=7 mT and $\nu_0$=30 MHz) a magnetic moment sensitivity of order $ 10^{-20} ~\mathrm{J} \, \perTesla$ or about $10^3 ~\mu_B ~\left( =9.27 \times 10^{-24} ~\mathrm{J} \, \perTesla \right)$.  The torque sensitivity is of order $ 10^{-22}$ J, much smaller than $k_B T \approx 4 \times 10^{-21}$ J at 300 K.  That torques smaller than the naive thermal limit are measurable is likely a consequence of the gyroscopic stabilization of the particle in the experiments\citep{Arita2013}.

\section{Conclusions and outlook} 

We have demonstrated a method of stabilizing levitated nanoplatelets by frequency-locking their rotation to an applied rf electric field, and we have used this technique to measure magnetic properties of graphene. The discrepancy between our data and theory may be addressed by improving characterization and quality of the samples in future experiments.  Improved characterization, most importantly confirmation of dimensions and mass, may be afforded by depositing nanoplatelets on a substrate\citep{Kuhlicke2015} after measurements on the levitated particles have been completed.  In addition to the uncertainty in nanoplatelet dimensions, it is also possible that impurities accumulate on the nanoplatelet from the solvent during the electrospray process. Impurities attached to the nanoplatelet could be removed by laser heating.  Unfortunately, we have found that heating also leads to discharge and loss of the particle, and future experiments will require a method to recharge the sample, such as a focused electron beam.  

We anticipate that our frequency-locking technique will enable a diverse range of measurements.  Most significantly, a gyroscopically-stabilized nanoplatelet could be used as a levitated mirror in the configuration of a traditional torsion balance or incorporated into cavity optomechanics experiments\citep{Williamson2016}, where the achievable torque sensitivities may be improved by several orders of magnitude\citep{Hoang2016b}.  Such a configuration may also be a sensitive probe of the flexural modes of the centrifugally tensioned nanoplatelet.  Our technique for sample translation and transfer\citep{Coppock2017} can be used to deliver samples to traps inside large magnets, allowing experiments to be extended into the quantum Hall regime, where torque measurements offer a unique probe of the \textit{equilibrium} state of two-dimensional systems\citep{Usher2009}\citep{Eisenstein1985}.  Rotational frequencies can also be significantly increased for graphene, possibly to 1 GHz, owing to its large tensile strength\footnote{The highest rate we have have yet achieved is $\nu_n\approx$100 MHz. We have observed the abrupt disappearance of samples spinning at rates between 20 and 100 MHz, which we have tentatively ascribed to decomposition from centrifugal forces. We have not yet undertaken systematic observations of this behavior.}.  Finally, our measurement technique may be applied to any material that can be prepared in a liquid suspension\citep{Nicolosi2013} and introduced into a vacuum system via electrospray techniques.

\section{Acknowledgements}

This work was supported by the Laboratory for Physical Sciences.

\bibliography{gyrolock_arxiv}

\end{document}